\documentclass{PoS}

\usepackage{graphicx}
\usepackage[utf8]{inputenc}
\usepackage[english]{babel}
\usepackage{amsmath,amssymb,color,amscd,subfigure}
\usepackage{amsfonts,amssymb,amsbsy}
\usepackage{booktabs}
\usepackage{calc}
\usepackage{cite}
\usepackage{amsthm,enumitem}
\usepackage{relsize}
\usepackage{multirow}
\usepackage{rotating}
\usepackage{bm}
\usepackage{url}
\usepackage{booktabs}
\usepackage{float}
\usepackage{graphicx}
\usepackage{multicol}
\usepackage{setspace}
\usepackage{ragged2e}

\allowdisplaybreaks
\providecommand*{\ord}{\ensuremath{\mathrm{O}}}
\providecommand*{\tr}{\ensuremath{\mathrm{Tr}}}
\providecommand*{\e}{\ensuremath{\mathrm{e}}}
\providecommand*{\I}{\ensuremath{\mathbf{1}}}
\providecommand*{\psibar}{\ensuremath{\overline{\psi}}}

\let\OLDthebibliography\thebibliography
\renewcommand\thebibliography[1]{
  \OLDthebibliography{#1}
  \setlength{\parskip}{-1.4pt}
  \setlength{\itemsep}{-1.1pt plus 0ex}
}

\title{Heavy-quark physics with a tmQCD valence action}

\ShortTitle{Heavy-quark physics with a tmQCD valence action}

\author{
\begin{minipage}[b]{0.4\linewidth}
\includegraphics[height=2.5\baselineskip]{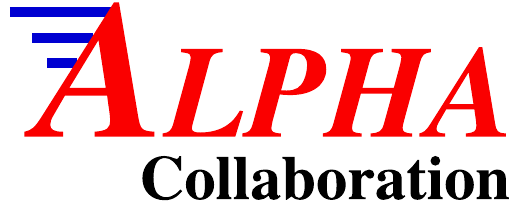}
\end{minipage}
\hfill
\begin{minipage}[b]{0.3\linewidth}
{\it
IFT-UAM/CSIC-18-100\\
FTUAM-18-22}
\end{minipage}
}

\author{\speaker{A.~Bussone}$\,\,^{a,b}$, 
S.~Chaves$\,^a$, 
G.~Herdo\'{i}za$\,^{a,b}$, 
C.~Pena$\,^{a,b}$,
D.~Preti$\,^{c}$,
J.\'{A}.~Romero$\,^{b}$, 
J.~Ugarrio$\,^{a,b}$\\
\llap{$^a$}{Department of Theoretical Physics, Universidad Aut\'{o}noma de Madrid, E-28049 Madrid, Spain}\\
\llap{$^b$}{Instituto de F\'{i}sica Te\'{o}rica UAM-CSIC, c/ Nicolás Cabrera 13-15, Universidad Aut\'{o}noma de Madrid, E-28049 Madrid, Spain}\\
\llap{$^c$}{INFN, Sezione di Torino
Via Pietro Giuria 1, I-10125 Turin, Italy}\\
E-mail:\,\email{andrea.bussone@uam.es}
}

\abstract{We introduce a mixed-action 
approach based on CLS ensembles, where a valence $N_f$=2+1+1 Twisted Mass QCD action is combined with the $N_f$=2+1 non-perturbatively $\ord(a)$-improved Wilson sea sector.
We show that for maximally twisted valence quarks, the automatic
$\ord (a)$-improvement of this set-up holds up to lattice artifacts coming from
sea quark mass ef\mbox{}fects.
Furthermore, we introduce a three-dimensional Gradient Flow smearing 
in order to tame the signal to noise ratio problem.}

\FullConference{The 36th Annual International Symposium on Lattice Field Theory\\
                 July 22–28, 2018\\
                 East Lansing, Michigan, USA}

\begin{document}

\section{Introduction}
\vspace{-0.3cm}
{\noindent}Heavy-quark physics is particularly challenging 
on the Lattice since 
various energy scales have to be accommodated simultaneously.
Finite volume corrections have to be negligible, resulting in the condition
$m_\pi L \gg 1$, and small discretization errors are required,
$a m_h \ll 1$, in order to have controlled cutof\mbox{}f ef\mbox{}fects.
An obvious (but expensive) 
way to decrease discretization ef\mbox{}fects
is to reduce the lattice spacing $a$,
but for $a\lesssim 0.05$ fm,
simulations are plagued by the topological freezing problem.
A complementary approach
is to employ the Symanzik improvement
programme \cite{Symanzik:1983dc}, which allows to systematically remove $\ord(a)$ ef\mbox{}fects 
in the observables of interest.\\
The CLS initiave has recently started a program of production of 
QCD configurations \cite{Bruno:2014jqa} with 
L\"uscher-Weisz tree-level $\ord(a^2)$-improved gauge action
and $N_f=2+1$ non-perturbatively $\ord(a)$-improved Wilson fermions.
The use of (tree-level improved) open boundary conditions (open BC)
for the fundamental degrees of freedom is of paramount importance in reaching 
small lattice spacing while avoiding the topology freezing by allowing a smooth
flow of the topological charge through the boundaries \cite{Luscher:2011kk}.\\
In addition to the reduction of the lattice spacing, it is also crucial to
implement an improvement programme when aiming at reliable lattice QCD
determinations of observables in the heavy-quark sector.
In \cite{Luscher:1996sc} it was elegantly explained how the $\ord(a)$ ef\mbox{}fects are
strictly connected to the hard breaking of chiral symmetry in the Wilson regularization
and the r\^{o}le of the symmetries in the improvement programme was uncovered.
Later, in \cite{Frezzotti:2003ni}, it was realized 
that physical observables are automatically O($a$)-improved
in the Twisted Mass (TM) regularization \cite{Frezzotti:2000nk},
once the standard mass is set to its critical value.\\
The question we are addressing in this work is whether the use of TM only
in the valence
sector of the theory still retains the automatic $\ord(a)$-improvement mechanism.
The na\"{i}ve expectation is that O($a$) ef\mbox{}fects arising from the sea regularization
could still contribute.
In the first part of this work we will briefly review the argument of $\ord(a)$
improvement for TM 
\cite{Frezzotti:2003ni} and combine it with
an extension of \cite{Bhattacharya:2005rb} to Wilson-like theories.\footnote{Regularizations for which the limit of vanishing
masses coincides with the Wilson regularization at zero mass.}
The extension of the latter work, which was originally formulated only for Wilson
fermions, is crucial in order to understand the contributions coming 
from sea and valence fermions to
$\ord(a)$ discretization ef\mbox{}fects.
In the second part we will discuss the use of the Gradient Flow technique 
\cite{Luscher:2013cpa}
restricted to an equal-time hyperplane in order to build
smeared interpolating operators.
Interpolating operators 
are important technical elements of 
heavy quark physics computations and the use of smearing
helps in getting signals at earlier Euclidean times.
\vspace{-0.3cm}
\section{$\ord(a)$-improvement of Wilson-like regularizations}
\label{sect:oa_impr}
\vspace{-0.3cm}
{\noindent}We set up the infinite volume lattice theory in the Euclidean space.
We restrict ourselves to an $\mathbf{SU}(N_c)$ gauge theory with $N_f$ non-degenerate
quarks in the fundamental representation.
Quarks and antiquarks $\psi, \psibar$ 
are taken as multiplets of $\mathbf{SU}(N_f)$.
We concentrate on the fermionic sector of the theory where the
unimproved lattice action is given 
by
\vspace{-0.3cm}
\begin{align}
\label{eq:w_action}
\mathrm{S} [U,\psibar,\psi] &= 
a^4 \sum_{x} \psibar(x) 
\left[ 
\frac{1}{2} \gamma_\mu \left( \nabla^+_\mu + \nabla^-_\mu \right) 
-\frac{a\, \mathbf{r}}{2} \nabla^+_\mu\nabla^-_\mu
+ \mathbf{m_{0}} + i \boldsymbol{\mu} \gamma_5 
\right]
\psi(x),
\end{align}
\vspace{-0.1cm}
with the standard and twisted bare quark-masses labeled by
$\mathbf{m_{0}}$ and $\boldsymbol{\mu}$, respectively, and $\mathbf{r}$ 
denotes the Wilson parameter.\footnote{Boldface symbols
 are used for matrices in flavor space.}
In the following we will refer to the 
standard
bare quark-mass as the subtracted one, given by
$\mathbf{m} = \mathbf{m_{0}}- m_{\mathrm{cr}}\I$.
Throughout this work we will only consider diagonal standard masses,
i.e., $\mathbf{m}=
{\rm diag}(m_{1}, m_{2}, \dots, m_{N_f})$,
and
$\boldsymbol{\mu}=\mu_a\, T^a$ for TM,
with $T^a$ 
generators
of $\mathbf{SU}(N_f)$,
while $\boldsymbol{\mu}=\mu_0\, \I + \mu_a\, T^a$ 
for the Osterwalder-Seiler (OS) case.
Notice that we are working in the so-called twisted basis, and the contact with the
Wilson formulation 
is easily established by setting the twisted-mass matrix to zero.\\
The discrete symmetries of the Wilson-like theory are the following \cite{Frezzotti:2003ni,Frezzotti:2004wz}:
\begin{itemize}[topsep=-2pt,itemsep=-2ex,partopsep=2ex,parsep=0ex]
\item $R_5 
\times [\mathbf{m}\rightarrow - \mathbf{m}] 
\times [\mathbf{r}\rightarrow - \mathbf{r}] 
\times [\boldsymbol{\mu}\rightarrow - \boldsymbol{\mu}]$, 
where $R_5$ is given by\footnote{In the continuum, 
$R_5$ is a symmetry of the integration measure for $N_f>1$
sea quarks \cite{Sint:2010eh}. When considering additional valence quarks, their
contribution to the integration measure can be exactly suppressed through the
inclusion of the corresponding ghost fields (see e.g.~\cite{Frezzotti:2004wz}).}
\vspace{-0.35cm}
\begin{align}
\psi(x) 
&\xrightarrow{R_5} \gamma_5 \psi(x), 
&\psibar(x) \xrightarrow{R_5} - \psibar(x) \gamma_5 .
\end{align}
\item $R_5 \times D$, where the $D$-transformation is given by
\vspace{-0.3cm}
\begin{align}
\psi(x) 
\xrightarrow{D} \e^{i3\pi / 2} \psi(-x),  
\quad \psibar(x) \xrightarrow{D} \e^{i3\pi/2} \psibar(-x), 
\quad U_\mu(x) \xrightarrow{D} U_\mu^\dagger(-x-\hat{\mu}).
\end{align}
\item $P \times [\boldsymbol{\mu} \rightarrow -\boldsymbol{\mu}]$, 
$T \times [\boldsymbol{\mu} \rightarrow -\boldsymbol{\mu}]$
and $C$, where $P$, $T$ and $C$ are respectively parity, 
time reversal and charge conjugation.
\end{itemize}
From the
symmetries listed above one can infer 
that also 
$P_5 
\times [\mathbf{m} \rightarrow -\mathbf{m}] 
\times [\mathbf{r} \rightarrow -\mathbf{r}]$ is
a symmetry of the action, where $P_5$ is given by the product $R_5\times P$.
\\
{\noindent}Let us consider for the moment $N_f$ massless Wilson fermions.
The theory possesses chiral symmetry and a convenient way to
write down the improvement is
given by, generalizing the notation
in \cite{Bhattacharya:2005rb},
$
\tr \,( T^a\, \Gamma \mathcal{O} ) \equiv
\psibar\, T^a\, \Gamma\, \psi 
= \tr \left\{ T^a\, \Gamma \left[ \psi \otimes \psibar \right]\right\}$,
where 
$\Gamma$ is an element of the Clif\mbox{}ford algebra.
Since we are considering Wilson-like theories the mass-independent improvement 
goes exactly as in \cite{Luscher:1996sc,Bhattacharya:2005rb},
while the massive one will be a slight modification of them.
We give, as examples,
the mass-dependent improvement of the Lagrangian masses and bilinears.
In \cite{Bhattacharya:2005rb}, 
in order to find the standard-mass dependence, a spurionic analysis was applied,
by considering the standard-mass matrix to transform in the adjoint
representation of $\mathbf{SU}(N_f)$.
We extend the spurionic analysis 
including 
the twisted-mass matrix, 
i.e., $i\,\gamma_5\,\boldsymbol{\mu} \rightarrow i\,\gamma_5\, U^\dagger \boldsymbol{\mu}\, U$.
The operators contributing to the massive O($a$) improvement are then found 
along the lines of \cite{Bhattacharya:2005rb}.
\vspace{-0.1cm}
\subsection{Lagrangian masses \& non-singlet bilinears}
\vspace{-0.1cm}
{\noindent}The improvement of the diagonal Lagrangian-masses is found to be,
taking into account the scalar (pseudoscalar) nature of the 
standard (twisted) masses, (see
\cite{Bhattacharya:2005rb} for any unexplained notation)
\vspace{-0.3cm}
\begin{align}
\nonumber
\widehat{m}_j = 
Z_m\, \bigg\{&
\left[
m_j + (r_m-1) \frac{\tr\,\mathbf{m}}{N_f}
\right]
+ a \bigg[
b_m\, m_j^2
+\widetilde{b}_m\, \mu_j^2
+ \overline{b}_m\, m_j\, \tr\,\mathbf{m}\\
\nonumber
&
+ (r_m\, d_m - b_m) \frac{\tr(\mathbf{m}^2)}{N_f}
+ (r_m\, \overline{d}_m - \overline{b}_m) \frac{(\tr\,\mathbf{m})^2}{N_f}
+ (r_m\, \widetilde{d}_m - \widetilde{b}_m) \frac{\tr(\boldsymbol{\mu}^2)}{N_f}
\bigg]
\bigg\},\\
\widehat{\mu}_j = 
Z_\mu\, \mu_j &
\left[ 
1
+ a 
\left(
b_\mu\,  m_j
+ \overline{b}_\mu\,  \tr\,\mathbf{m}
\right)
\right].
\end{align}
As another example of the improvement out of the chiral limit we give 
the expressions for non-singlet bilinears 
\vspace{-0.3cm}
\begin{align}
\nonumber
\widehat{\tr (T^a\, \Gamma\, \mathcal{O} )}  = 
Z_{\mathcal{O}_{\Gamma}} \bigg\{ &
\tr (T^a\, \Gamma\, \mathcal{O} )^I
\left[ 
1 + a\overline{b}_{\mathcal{O}_{\Gamma}}\, \tr\, \mathbf{m}
\right]
+a\frac{b_{\mathcal{O}_{\Gamma}}}{2}\,
\tr \left(
\{T^a,\mathbf{m}\} \Gamma\,\mathcal{O}
\right)\\
\nonumber
& 
+af_{\mathcal{O}_{\Gamma}}\,
\tr \left( T^a\mathbf{m}\right)
\tr \left(
 \Gamma\,\mathcal{O}
\right)
+a \widetilde{b}_{\mathcal{O}_{\Gamma\gamma_5} }
\tr \left(
T^a \, i\boldsymbol{\mu}\, \Gamma\,\gamma_5\,\mathcal{O}
\right)\\
\label{eq:nonsingletimpr}
&
+a \widetilde{f}_{\mathcal{O}_{\Gamma\gamma_5}}
\tr \left( T^a\, i \boldsymbol{\mu}\right)
\tr \left(
 \Gamma\,\gamma_5\,\mathcal{O}
\right)
+
a \widetilde{\overline{b}}_{\mathcal{O}\gamma_5}
\tr\, i\boldsymbol{\mu}\,
\tr (T^a\,\Gamma\,\gamma_5\,\mathcal{O})
\bigg\}.
\end{align}
Additional terms appear with respect to \cite{Bhattacharya:2005rb}, 
some of which are 
absent in TM, while present in the OS case.
\begin{figure}[!b]
\vspace{-0.3cm}
\begin{center}
\includegraphics[scale=0.52]{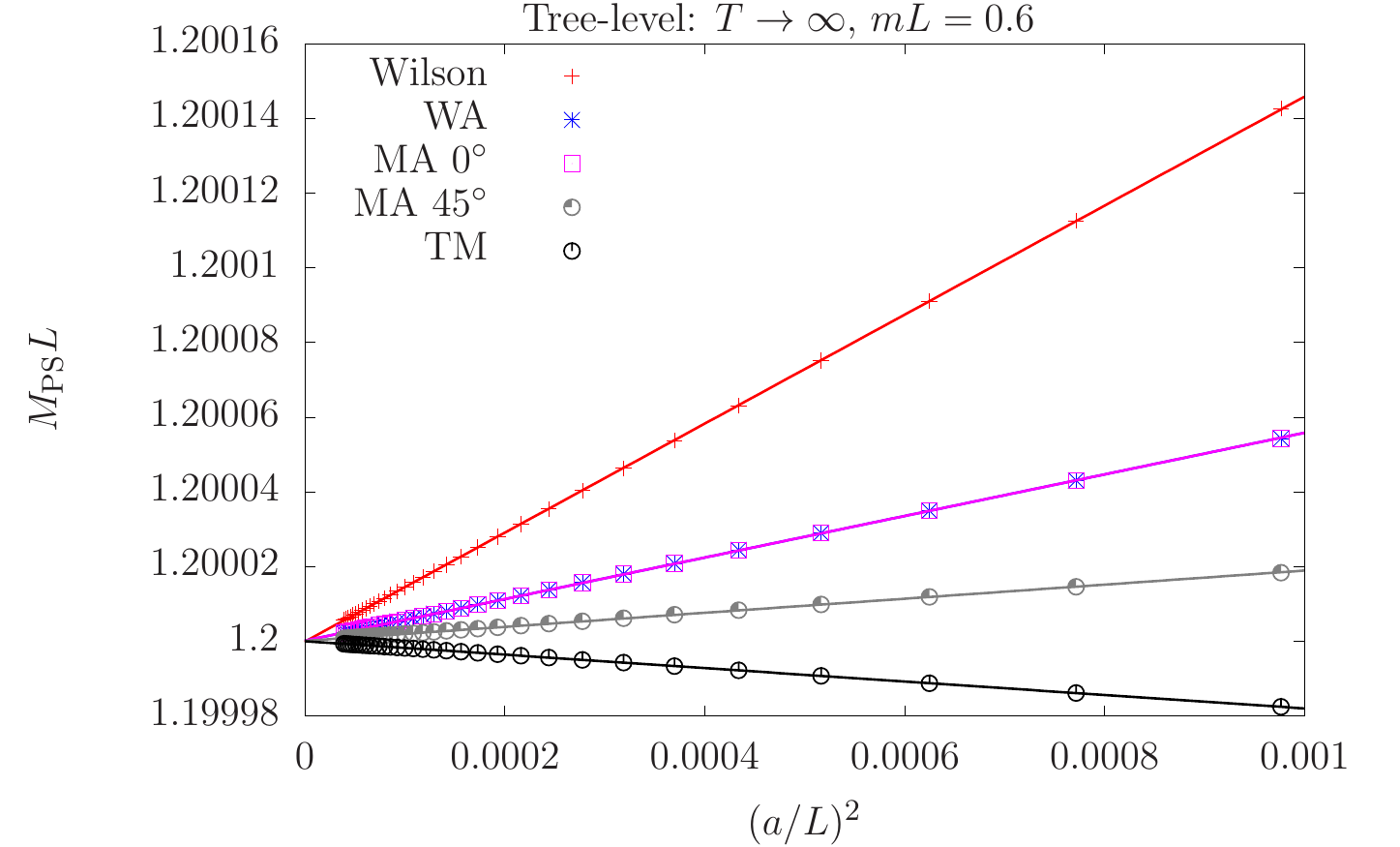}
\hspace{-0.5cm}
\includegraphics[scale=0.52]{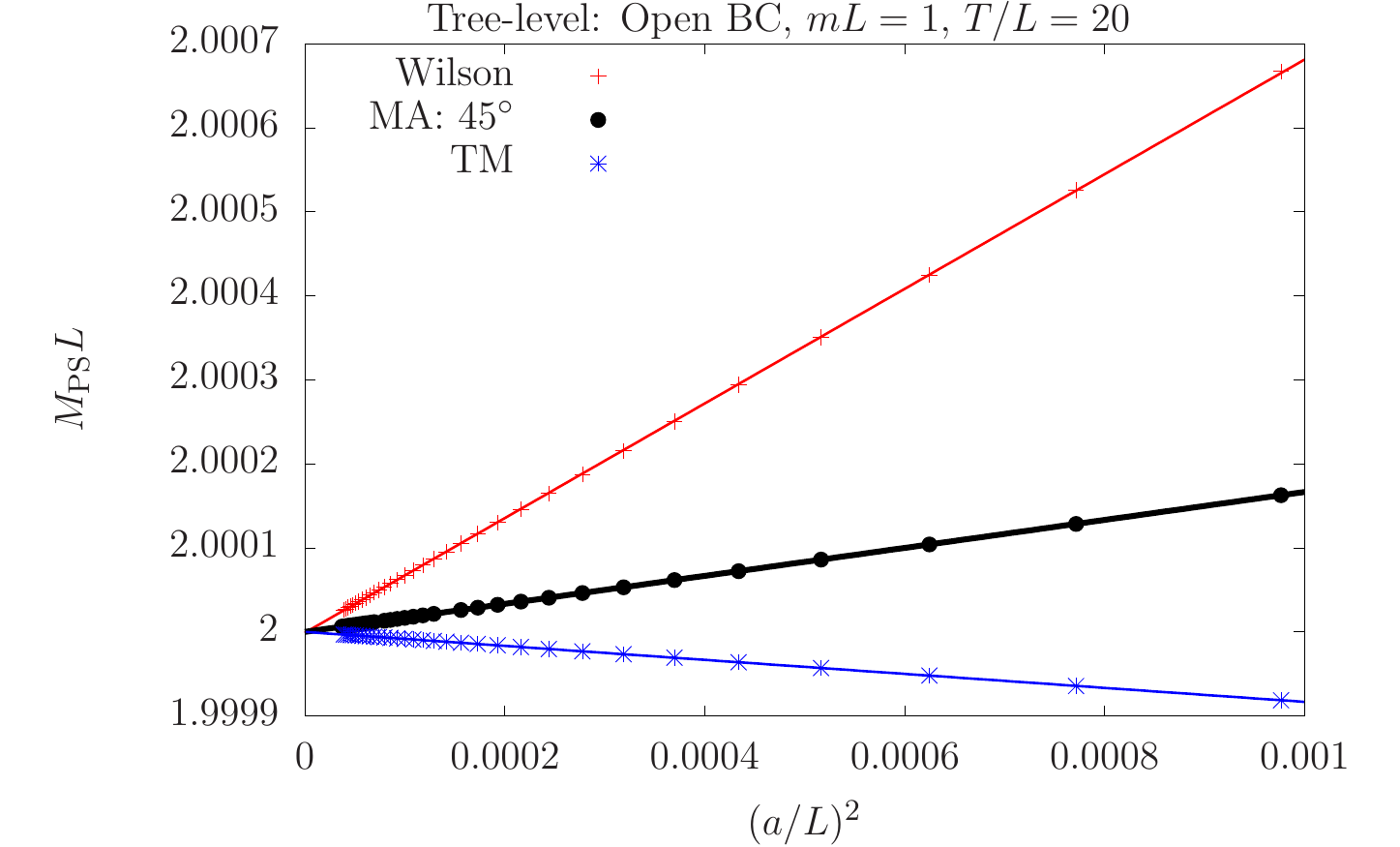}
\caption{Left panel: Tree-level scaling of the 
mass-degenerate pseudoscalar meson mass in the infinite time extent limit.
See \cite{Cichy:2008gk} for a similar study.
The Wilson data show O($a^2$) scaling once the standard quark masses are improved.
In the other cases 
no improvement of the masses is performed 
and the O($a^2$) scaling is clearly visible.
WA and MA (at vanishing angle) results are equivalent, since they are
related by a symmetry.  
Right panel: Similar plot as in the left panel but 
in the open BC set-up. 
The formulae employed were taken from \cite{Luscher:1996vw,Frezzotti:2001ea}.
As before, the Wilson case has to be improved before getting the expected
O($a^2$) scaling.}
\label{fig:improvement}
\end{center}
\end{figure}
\vspace{-.3cm}
\subsection{On-shell automatic $\ord(a)$-improvement}
\vspace{-0.2cm}
{\noindent}In this section we recall the proof of 
automatic $\ord(a)$-improvement given in \cite{Frezzotti:2003ni}.
Once the fermionic action and the bilinears are improved, one
can be convinced that all the operators
entering at $\ord(a)$ in
physical Green's functions
are odd under sign exchange of the Wilson parameter.
By considering the Wilson Averaging (WA) technique \cite{Frezzotti:2003ni} 
and the symmetries given in Sect.~\ref{sect:oa_impr} 
we obtain the improvement of Wilson-like theories.\footnote{The inclusion
of $c_{\rm SW}$ does not alter the conclusions in this section, the 
associated operator is odd under $r\rightarrow -r$.}
A higher power for $a^n$ corrections 
is found once the WA is applied on the $n$-point multiplicatively renormalizable
correlation function $\Phi$,
\vspace{-0.3cm}
\begin{align}
\frac{1}{2}
\left[
\langle \Phi \rangle \Bigr|_{\substack{\mathbf{r} \\ \mathbf{m}\\ \boldsymbol{\mu}}}
+
\langle \Phi \rangle \Bigr|_{\substack{-\mathbf{r} \\ \phantom{-}\mathbf{m}\\ \phantom{-}\boldsymbol{\mu}}}
\right]
= 
\langle \Phi_0 \rangle +\ord(a^2).
\end{align}
Where the symbol $\langle\dots\rangle\bigr|_{\substack{\mathbf{r}, \dots}}$ refers to the expectation value calculated in the Lattice theory with parameters
$\mathbf{r}, \dots$, and $\Phi_0$ is the continuum counterpart 
of the lattice correlator $\Phi$.
By using the $P_5 
\times [\mathbf{m} \rightarrow -\mathbf{m}] 
\times [\mathbf{r} \rightarrow -\mathbf{r}]$ 
symmetry one gets to the Mass Average (MA) \cite{Frezzotti:2003ni}
\vspace{-0.3cm}
\begin{align}
\frac{1}{2}
\left[
\langle \Phi \rangle \Bigr|_{\substack{\mathbf{r} \\ \mathbf{m}\\ \boldsymbol{\mu}}}
+
\eta^P \eta^{R_5}
\langle \Phi \rangle \Bigr|_{\substack{\phantom{-}\mathbf{r} \\ -\mathbf{m}\\ \phantom{-}\boldsymbol{\mu} }}
\right]
= 
\langle \Phi_0 \rangle +\ord(a^2),
\end{align}
where we have denoted the parity and $R_5$-parity of the correlator with the
corresponding $\eta$'s.
Now if we consider
vanishing standard masses\footnote{It is suf\mbox{}ficient 
that the masses vanish up to O($a$) ef\mbox{}fects.} 
we obtain 
the automatic on-shell O($a$)-improvement for physical
observables.
In Fig.~\ref{fig:improvement} we show the mechanism of O($a$)-improvement
at tree-level in the infinite time extent limit 
and with Dirichlet boundary conditions in the time direction.\footnote{The latter are
the fermionic boundary conditions used in CLS open BC set-up.}
\vspace{-0.3cm}
\section{Mixed action improvement}
\vspace{-0.3cm}
{\noindent}We are now ready to set up the mixed action theory with Wilson fermions in the sea and
TM in the valence sector.
In \cite{Luscher:1976ms}
a self-adjoint, strictly positive transfer matrix
was explicitly constructed for the 
the Wilson regularization ($r=1$)\footnote{In 
\cite{Frezzotti:2003ni} the theory with $r=-1$ 
was proven to possess site reflection positivity.},
and later on the proof was extended to the
TM regularization \cite{Frezzotti:2001ea,Shindler:2007vp}.
Mixed actions
are known to be non-unitary,\footnote{A valence and sea matching, such that the correct
continuum limit theory is obtained, 
can be performed, see \cite{Herdoiza:2017bcc,Bussone:tobe1} for 
an example related to this work.} since the valence 
determinant has to be suppressed by ghosts \cite{Morel:1987xk}.
The symmetries of the valence action are easily extended to the ghost sector \cite{Frezzotti:2004wz,Golterman:2005ie}, but its inspection is not strictly necessary as long
as we are interested in correlators made of valence quarks only.
By using the valence WA and MA 
one discovers 
that O($a$)-ef\mbox{}fects proportional to the trace of the sea-quark mass matrix
  are not eliminated.
As an example of these findings we give here the comparison between
the quark-mass lattice artifacts in the Wilson, TM and mixed action specified above
\vspace{-0.3cm}
\begin{align}
\nonumber
\text{Wilson}: \,\,&\widehat{m}_{ij} = 
\frac{Z_A}{Z_P}\,
 m_{ij} 
\left[
1+ a (\widetilde{b}'_A - \widetilde{b}'_P)\, m_{ij}
+ a (\overline{b}'_A - \overline{b}'_P) \tr\,\mathbf{m}
\right] 
+\ord(a^2),
&
\text{[PCAC mass]}
\\
\nonumber
\text{TM}: \,\,&\widehat{\mu}_j = 
\frac{1}{Z_P}\, \mu_j 
+\ord(a^2),
&
\text{[Lagrangian mass]}\\
\text{Mixed action}: \,\,&\widehat{\mu}_j = \frac{1}{Z_P} \,
\mu_j \left( 1+
a  \overline{b}_\mu\, \tr\,\mathbf{m}
\right)
 + \ord(a^2),
 &
\text{[Lagrangian mass]}
\end{align}
where $\overline{b}'_A  = \overline{b}'_P = \overline{b}_\mu = \ord(g_0^4)$ start at two-loop in perturbation theory and 
$\widetilde{b}'_A - \widetilde{b}'_P = -0.0012g_0^2 + \ord(g_0^4)$ \cite{Taniguchi:1998pf}.
It is worth noting that the mixed action is free from valence O($a\mu$) ef\mbox{}fects
while cutof\mbox{}f ef\mbox{}fects proportional to the trace of the sea quark mass, 
$\tr\, \mathbf{m}$, can contribute.
Preliminary results on 
the charm mass scaling in $a$, obtained with 
the TM action in \cite{Pena:2004gb}, are presented
in \cite{Bussone:tobe2}.
\vspace{-0.3cm}
\section{Smearing of interpolating operators}
\vspace{-0.3cm}
{\noindent}A signal-to-noise ratio problem arises in the calculation of 
correlators when there are states contributing
to the variance with energies smaller
than twice the ground state energy. 
Its manifestation can be very severe, especially at large 
source-sink time separations, since the degradation 
of the signal is exponential
\cite{Parisi:1983ae,Lepage:1989hd}.
A way to tame the problem is to employ an interpolating operator that has 
better overlap with the ground
state, thus enabling the extraction of relevant quantities at earlier time separations.
The smearing corresponds to employing non-local operators in space, 
that attempt to replicate the extended nature of hadronic states.
The smearing of interpolating operators is a standard technique, with
  the Wuppertal smearing being one 
of the most prominent examples \cite{Alexandrou:1990dq}.
Here we introduce the Gradient Flow smearing \cite{Luscher:2013cpa}
in an equal-time hyperplane
(3D-GF).
The flow equations are a coupled system of dif\mbox{}ferential equations
\vspace{-0.3cm}
\begin{align}
\begin{cases}
\partial_t\, B_j(t,x) = D_k G_{k j}(t,x)\\
\partial_t\, \chi(t,x) = \nabla^2\, \chi(t,x)\\
\partial_t\, \overline{\chi}(t,x) = \overline{\chi}(t,x)\, \overleftarrow{\nabla^2 }
\end{cases}
\text{with initial conditions:}\,\,\,\,\,\,\,
\begin{cases}
B_j(t,x)\big|_{t=0} = A_j(x)\\
\chi(t,x)\big|_{t=0} = \psi(x)\\
\overline{\chi}(t,x)\big|_{t=0} = \overline{\psi}(x)
\end{cases},
\end{align}
where $\nabla^2 = D_j\, D_j$, with $D_j = \partial_j + B_j$.
Note that compared to \cite{Luscher:2013cpa} the temporal components of the gauge field are not evolved.
Furthermore, in contrast to Wuppertal smearing \cite{Alexandrou:1990dq} the evolution of
the fermionic and gauge degrees of freedom is done synchronously.
The 3D-GF is still a Gaussian smearing with tree-level radius squared given by $R^2 = 2D\, t$, and $D=3$.
Upon discretization of the flow equations one can use dif\mbox{}ferent 
integrators \cite{Luscher:2013cpa}.
In the case in which one fixes the background gauge field, for example (but not necessarily) at time $t=0$, the 3D-GF coincides with Wuppertal smearing, by further choosing the Euler integrator.
In \cite{Luscher:2013cpa} the GF theory was proven to have
good renormalization properties, in particular
correlators of an arbitrary number of evolved gauge fields are finite 
and do not require further renormalization once the boundary (QCD) is renormalized.
Na\"{i}vely the same can be expected in the 3D-GF framework if only correlators of gauge
fields of type $B_j$ are considered, but further studies on the
subject are required.\\
In \cite{Shindler:2013bia} it was noted that no extra O($a$)-ef\mbox{}fects arise when considering the GF 
of a TM regularization with a degenerate doublet of quarks at maximal twist.
A generalization of the arguments given in Sect.~\ref{sect:oa_impr} allows to prove the absence of $\ord(a)$
ef\mbox{}fects in both the GF and the 3D-GF when considering non-degenerate masses.
\vspace{-0.2cm}
\begin{figure}[!t]
\includegraphics[scale=0.55]{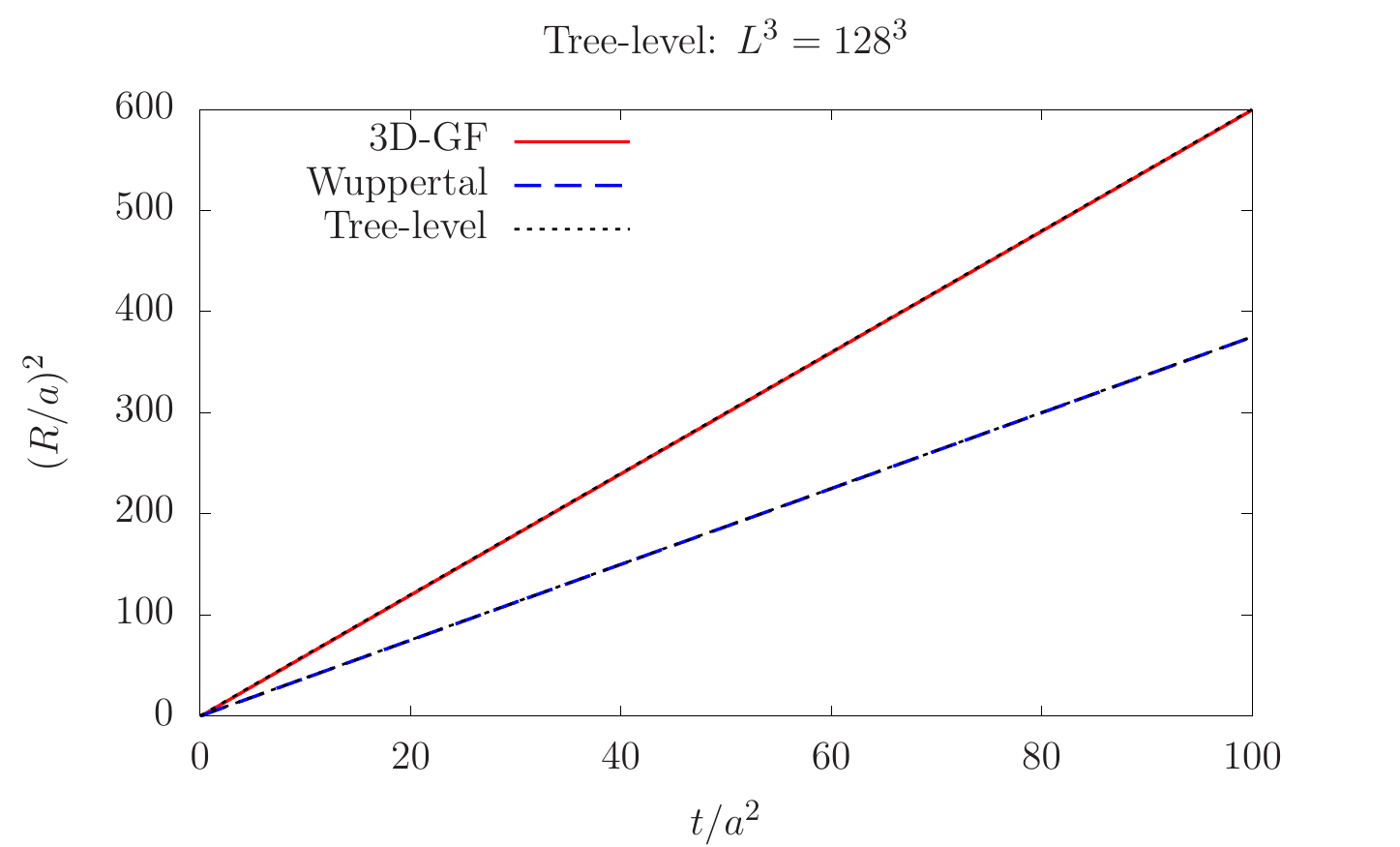}
\hspace{-.5cm}
\includegraphics[scale=0.55]{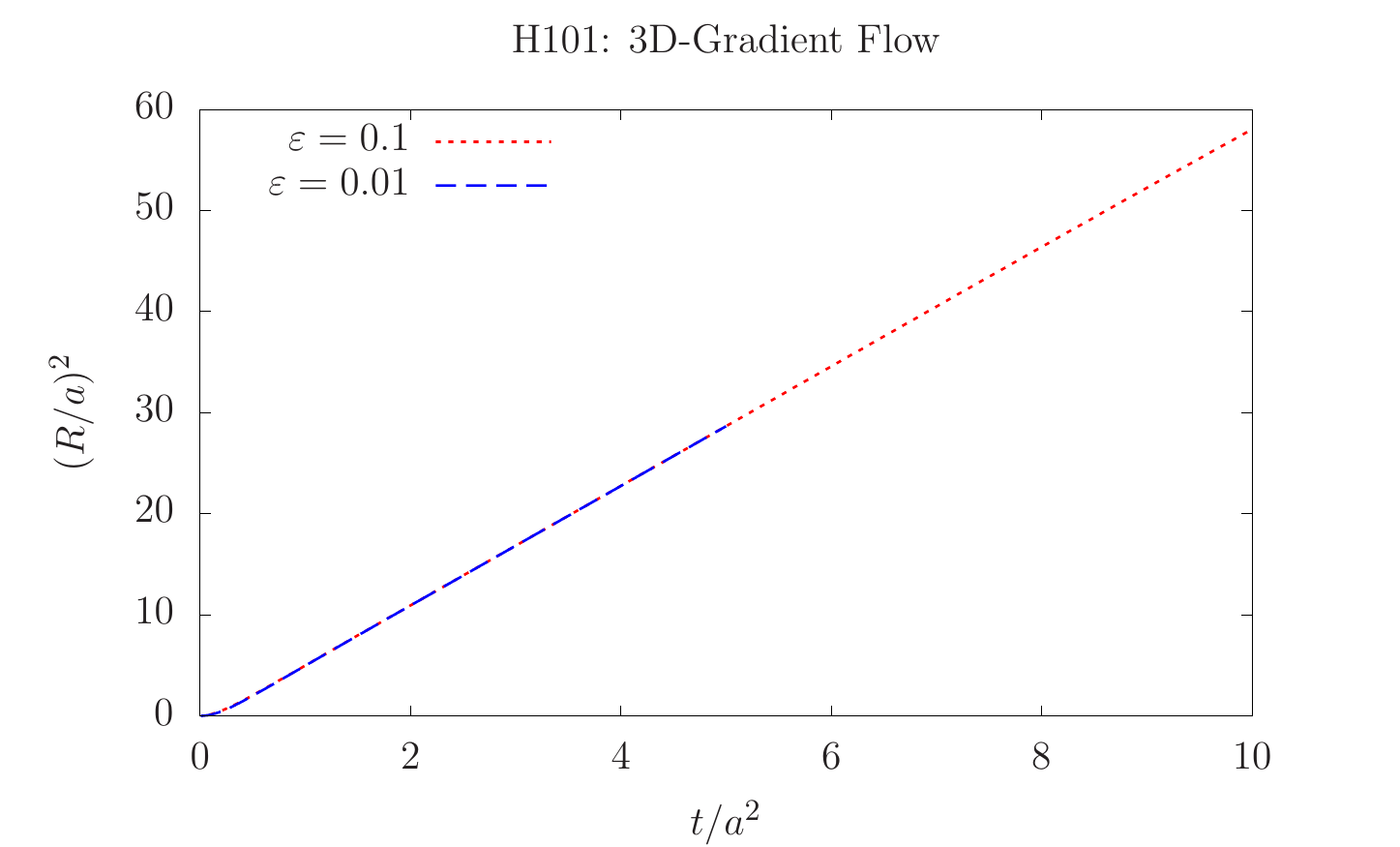}
\caption{Left panel: smearing radius squared in the tree level theory as a function
of the flow time. 
GF and Wuppertal agree with the theoretical expectations.
Right panel: smearing radius squared of one configuration of the H101 ensemble
($m_\pi = m_K = 420$ MeV, $a=0.086$ fm)
as a function of the flow time.}
\label{fig:radii}
\end{figure}
\begin{figure}[!t]
\begin{center}
\includegraphics[scale=0.65]{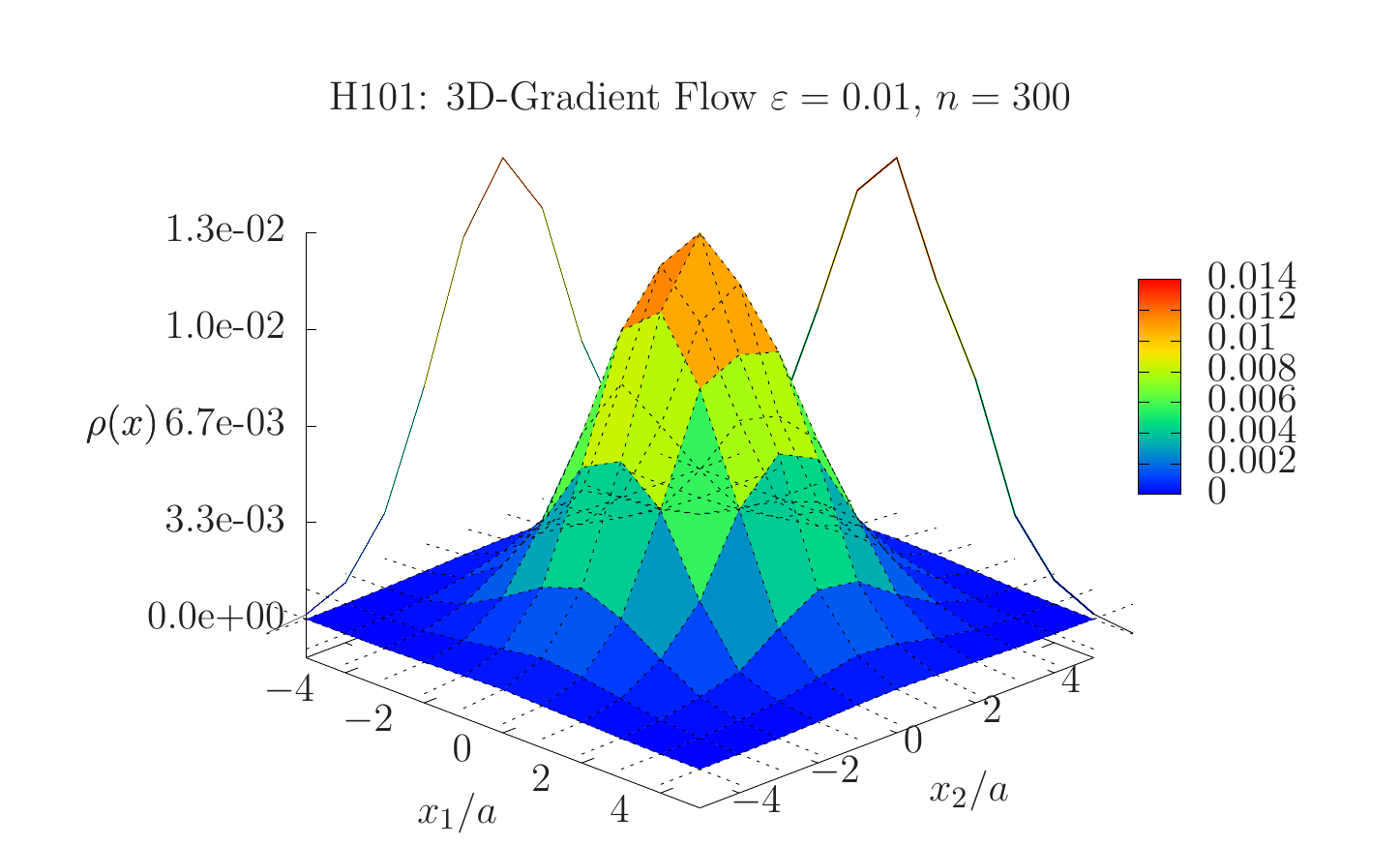}
\caption{$\rho(x)$ defined as $\rho(t;x) = |\chi(t;x)|^2 / \sum_x |\chi(t;x)|^2$
at $t/a^2=3$, with $x_0/a=48$ and $x_3/a=0$.}
\label{fig:adj_flow}
\end{center}
\end{figure}
\subsection{Tree-level results}
\vspace{-0.25cm}
{\noindent}In the free theory, by starting from a $\delta$-function in position and color space,
the smearing produces a Gaussian shape with radius squared $R^2$ given by:
$6t$ for 3D-GF and $6t/(1+6\varepsilon)$ for Wuppertal,
where $t/a^2 = n\, \varepsilon$, with $n$ iteration number and $\varepsilon$ 
step size.
The results in the free theory are shown in the left panel of Fig.~\ref{fig:radii}, 
where a perfect matching with the expectations is found.
In the right panel of Fig.~\ref{fig:radii} we show the 3D-GF performed
on one configuration
of the H101 ensemble \cite{Bruno:2014jqa}, with two dif\mbox{}ferent 
values of $\varepsilon$, which are in complete agreement.
In Fig.~\ref{fig:adj_flow} we show the result of the application of the 3D adjoint
flow on a $\delta$-function source.
The results are obtained through an extension of the code
given in \cite{Luscher:2012av}.
\vspace{-.4cm}
\section{Conclusions}
\vspace{-0.4cm}
{\noindent}We have presented 
an analysis of the $\ord(a)$-improvement 
by extending the work done in \cite{Bhattacharya:2005rb}
to Wilson-like theories.
We have analyzed the improvement in the case of a mixed action consisting of a TM regularization of the valence 
and a Wilson regularization of the sea sector.
We have found that 
the automatic $\ord(a)$-improvement is valid up to cutof\mbox{}f
ef\mbox{}fects coming from sea quark masses
and applied the Symanzik improvement programme to identify the relevant
operators.
Finally, we have presented the 3D Gradient Flow
as a tool to fight the signal to noise ratio problem.\\
{\noindent}
{\bf Acknowledgments:}  A.B.~wishes to thank P.~Dimopoulos and R.~Frezzotti for useful discussions
on the Twisted Mass regularization. We acknowledge PRACE for awarding us access to MareNostrum at the Barcelona
Supercomputing Center (BSC), Spain. We acknowledge the use of the Hydra cluster at the Instituto
de F\'{i}sica Te\'{o}rica (IFT) were some part of the calculations were done.
We acknowledge support through the Spanish MINECO project
FPA2015-68541-P, the Centro de Excelencia Severo Ochoa Programme SEV-2016-
0597 and the Ram\'{o}n y Cajal Programme RYC-2012-10819.
We are grateful to CLS members for producing the gauge configuration ensembles used in this study.

\end{document}